\documentstyle[11pt,newpasp,twoside,epsf]{article}
\def\edcomment#1{\iffalse\marginpar{\raggedright\sl#1\/}\else\relax\fi}
\reversemarginpar

\begin{document}
\title{A Deep, Wide-Field H$\alpha$ Survey of Nearby Clusters of Galaxies}
 \author{Shoko Sakai$^1$, Robert Kennicutt$^2$, Chris Moss$^3$}
\affil{$^1$UCLA \& KPNO/NOAO, $^2$Steward Observatory, $^3$Liverpool John Moore University}

\begin{abstract}
We present a progress report on an ongoing H$\alpha$ 
imaging survey of nearby clusters of galaxies.
Four clusters have been surveyed to date: A1367, A1656, A347 and A569.
A preliminary comparison of H$\alpha$ luminosity functions obtained 
from our imaging survey with those from the prism survey reveals a 
significant level of incompleteness in the latter.  This in turn
is due to a combination of insensitivity to low-luminosity emission-line
galaxies and to brighter galaxies with weak extended H$\alpha$
emission.   The survey has also revealed a unique population of
clustered dwarf emission-line objects which may be the results
of recent tidal encounters between larger gas-rich galaxies.

\end{abstract}

\section{Introduction}

With the Mosaic CCD Imager on the 0.92m telescope at KPNO,
we are able to cover
an 1$^{\circ}$ field in a single observation, and reach
limiting fluxes corresponding to star formation rates
(SFRs) of about 0.1 M$_\odot$/yr for extended galaxies 
and $\sim$0.01 M$_\odot$/yr for compact emission-line galaxies.
The broad goals of the survey are to construct a complete
inventory of star forming galaxies in a well-defined volume
limited sample, to characterize the demographics of the local
star forming population, and to quantify the completeness of
other methods for measuring the local SFR density.  We are 
also carrying out a parallel survey of a local field sample,
so we also will be able to study the influences of cluster
environment on the SFRs.

\section{Luminosity Functions}
We have observed four clusters so far during an observing run in
February 1999. These clusters have been observed
previously by Moss, Whittle, \& Pesce (1998: MWP98) using an objective prism technique.
In Figure~1, we show a flux distribution of H$\alpha$-emitting galaxies
in A1367.  Also shown by shaded histograms are the distributions of
the galaxies detected by MWP98.
Although the analysis is still in the preliminary stage, 
the comparison between our data and those from MWP98's prism survey suggests
that $\sim 50-60$\% of H$\alpha$ flux is missing from the latter.
This is due to a combination of insensitivity of the prism survey 
to low-luminosity emission-line
galaxies and to brighter galaxies with weak extended H$\alpha$ emission.   
In addition, we note that we have detected a large number
H$\alpha$-emitting galaxies without any velocity data.
Since membership is uncertain for these objects,
we have not included them in any of the analysis.
However, it is evident that the imaging survey has
revealed the incompleteness in the detection of H$\alpha$ fluxes
in the previous prism surveys.

\section{Tidally-Induced Dwarf Galaxies?}
While visually searching for emission-line galaxies in the central field of
Abell~1367, we found a particularly interesting field around N3860,
CGCG~97-114 and CGCG~97-125, scattered with emission-line dwarf knots.
Each ``knot'' is only a few arcsec in diameter, however, some 
are clustered in small groups that likely are par of dwarf galaxies.
With the present data, we cannot answer
as to what triggered such widespread star formation in this
particular region of the cluster.
We have obtained spectra for some H$\alpha$-emitting objects and 
they all coincide with those of CGCG97-114 and 97-125 ($v \sim 8200$ km/s), 
confirming that they are associated with the larger galaxies in this group.
(NGC~3860 is not in this group; its velocity is $\sim 6000$km/s).
The star formation rates for these objects, calculated using 
Kennicutt (1983), vary from $0.02$ up to $1.2$M$_{\odot}$/year.

This research was supported by NSF Grant AST-9900789.
SS was supported by the NASA LTSA program, NAS7-1260.

\begin{figure}
\plotfiddle{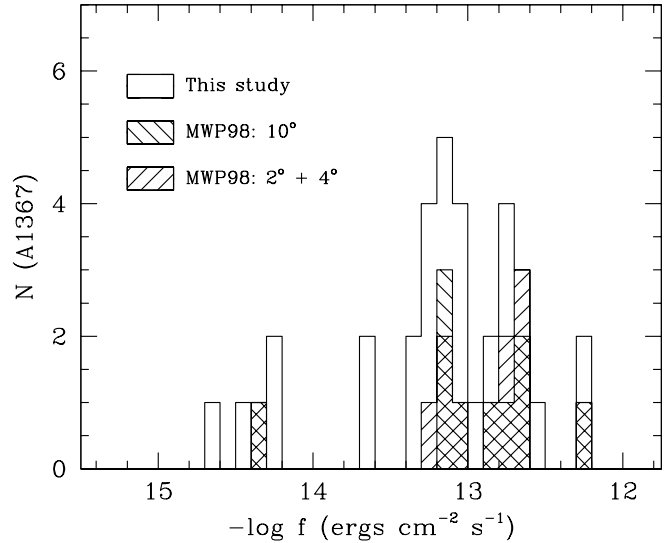}{2.75in}{0}{50}{50}{-150}{-130}
\caption{A H$\alpha$ flux distribution for galaxies in A1367, detected
in our imaging survey. It is compared to those from the prism survey of
MWP98.}
\end{figure}

\end{document}